\documentclass[11pt,twoside]{article}
\usepackage{asp2006}
\usepackage{epsf}
\usepackage{lscape}
\begin{document}
\title{
Galaxy Evolution in Dense Environments: Properties of Interacting 
Galaxies in the A901/02 Supercluster
}

\vspace{-0.2cm} \author{A.L. Heiderman (UT Austin), S. Jogee (UT
Austin), D.J. Bacon (Portsmouth), M.L. Balogh (Waterloo), M. Barden
(Innsbruck), F.D. Barazza (EPFL), E.F. Bell (MPIA), A. B\"ohm (AIP),
J.A.R. Caldwell (UT Austin), M.E. Gray (Nottingham), B. H\"au\ss ler
(Nottingham), C. Heymans (UBC, IAP), K. Jahnke (MPIA), E. van Kampen
(Innsbruck), S. Koposov (MPIA), K. Lane (Nottingham), D.H. McIntosh
(UMass), K. Meisenheimer (MPIA), C. Y. Peng (NRC-HIA), H.-W. Rix
(MPIA), S.F. Sanchez (CAHA), R. Somerville (MPIA), A.N. Taylor (SUPA),
L. Wisotzki (AIP), C. Wolf (Oxford), \& X. Zheng (PMO)}
\vspace{-0.4cm}

\begin{abstract}
We present a study of galaxies in the STAGES survey of the Abell
901/902 supercluster at $z\sim$~0.165, based on $HST$ ACS F606W,
COMBO-17, $Spitzer$ 24$\micron$, XMM-Newton X-ray, and gravitational
lensing maps.  We characterize galaxies with strong,
externally-triggered morphological distortions and normal, relatively
undisturbed galaxies, using visual classification and quantitative CAS
parameters.  We compare normal and distorted galaxies in terms of
their frequency, distribution within the cluster, star formation
properties, and relationship to dark matter (DM) or surface mass
density, and intra-cluster medium (ICM) density.  We report here our
preliminary results.
\end{abstract}

\vspace{-1cm}
\section{Introduction and Methodology}
In order to place constraints on galaxy evolution in cluster
environments, we present a study based on the STAGES survey of the
Abell 901/902 supercluster (Gray et al. 2008) at $z\sim$~0.165.  The
dataset includes high resolution ($0.1\arcsec$, corresponding to
$\sim$280\footnote{We assume in this paper a flat cosmology with
$\Omega_M = 1 - \Omega_{\Lambda} = 0.3$ and $H_{\rm 0}$
=70~km~s$^{-1}$~Mpc$^{-1}$.} pc) $HST$ ACS F606W images, star
formation rates (SFRs) from $Spitzer$ 24$\micron$ (Bell et al. 2007)
data, XMM-Newton X-ray data, and gravitational lensing maps (Heymans
et al. 2008), along with accurate spectrophotometric redshifts from
COMBO-17 (Wolf et al. 2004) and stellar masses (Borch et al. 2006).

Using the CAS code (Conselice et al. 2000), the concentration (C),
asymmetry (A), and clumpiness (S) parameters were derived from $HST$
ACS F606W images.  Since the CAS merger criteria (A$>$S and A$>$0.35)
tend to capture only a fraction of interacting/merging galaxies (e.g.,
Conselice 2006; Jogee et al. 2008a,b), we also resort to visual
classification.  We classify galaxies into three distinct visual
classes: (1) Galaxies with {\it externally-triggered} distortions:
these are distortions that arise from tidal interactions or mergers
and include features, such as double or multiple nuclei inside a
common body, tidal tails, arcs, shells, ripples, or tidal debris in
body of galaxy, warps, and offset rings.  (2) Galaxies with {\it
internally-triggered} asymmetries (classified as Irr-1):
internally-triggered asymmetries are due to stochastic star-forming
regions or the low ratio of ordered to random motions, common in
irregular galaxies.  These asymmetries tend to be correlated on scales
of a few hundred parsecs, rather than on scales of a few kpc.  (3)
Relatively {\it undistorted symmetric} galaxies (classified as
Normal).

Visual classification shows that 3.3$\pm$1\% of galaxies brighter
than M$_{\rm V} \leq$ $-$18 are strongly distorted. The fraction of
strongly distorted galaxies with M$_{\star} > 2.5\times10^{10} \rm
M_{\sun}$ is $\sim$1\%.  Most (74\%) of strongly distorted
galaxies lie on the blue cloud (Fig. 1, right).  The distortion fraction
in A901/02 is lower than in the field (7\% to 9\%) for massive or
bright (Jogee et al 2008a,b; Lotz et al. 2008) galaxies at similar
redshifts.

Most strongly distorted galaxies on  A901/902 lie outside the cluster cores, 
and avoid the peaks in DM surface mass density ($\kappa \geq$ 0.1) and ICM 
density (Fig. 1, left). These results are consistent with high galaxy velocity
dispersions in the core being unfavorable to mergers and strong tidal
interactions.  

\vspace{-0.2 cm} 
\acknowledgements AH and SJ acknowledge support from
NSF grant AST-0607748 amd LTSA grant NAG5-13063.

\vspace{-0.3cm}
\setcounter{figure}{0}
\begin{figure}[!ht]
\plottwo{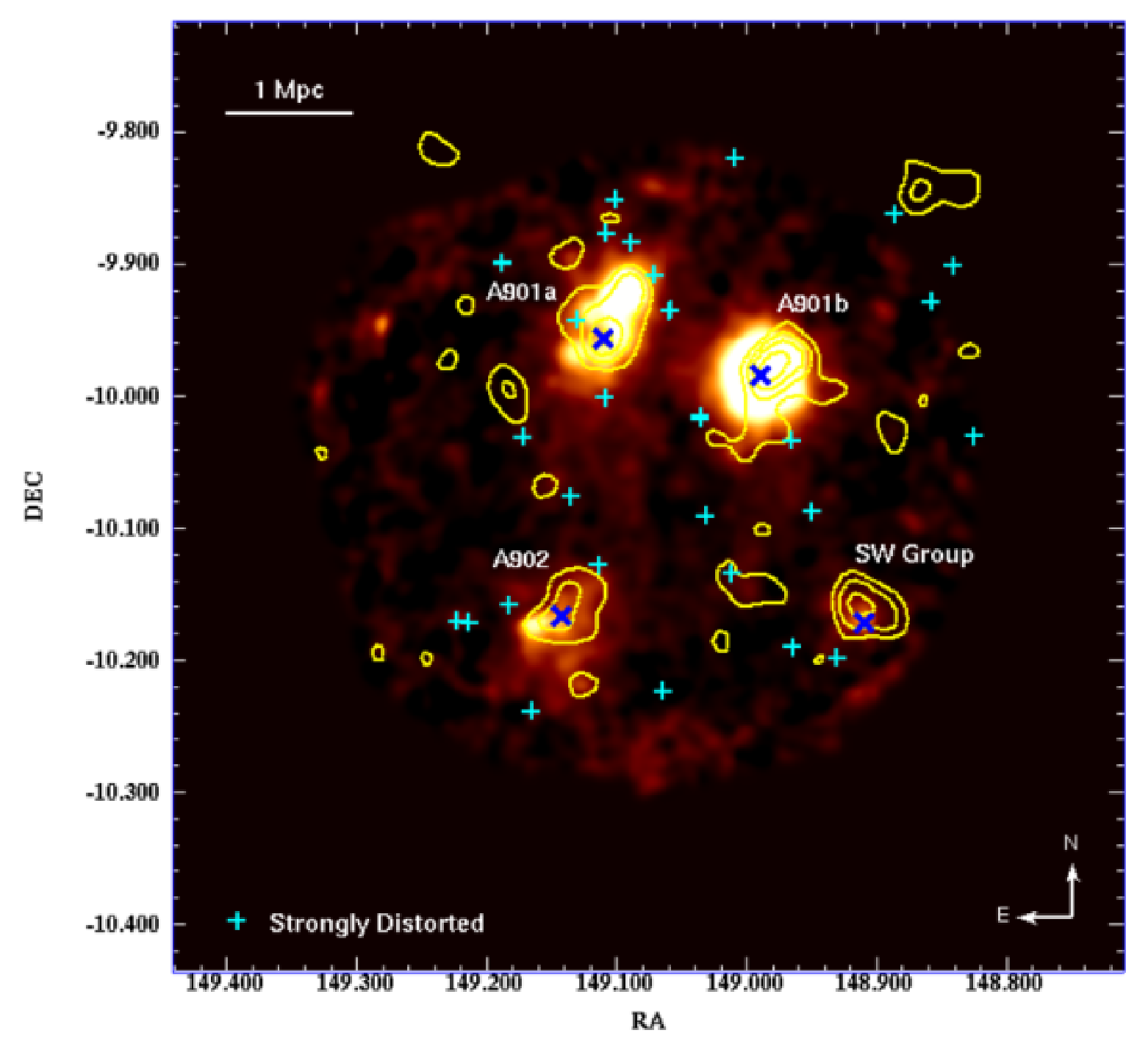}{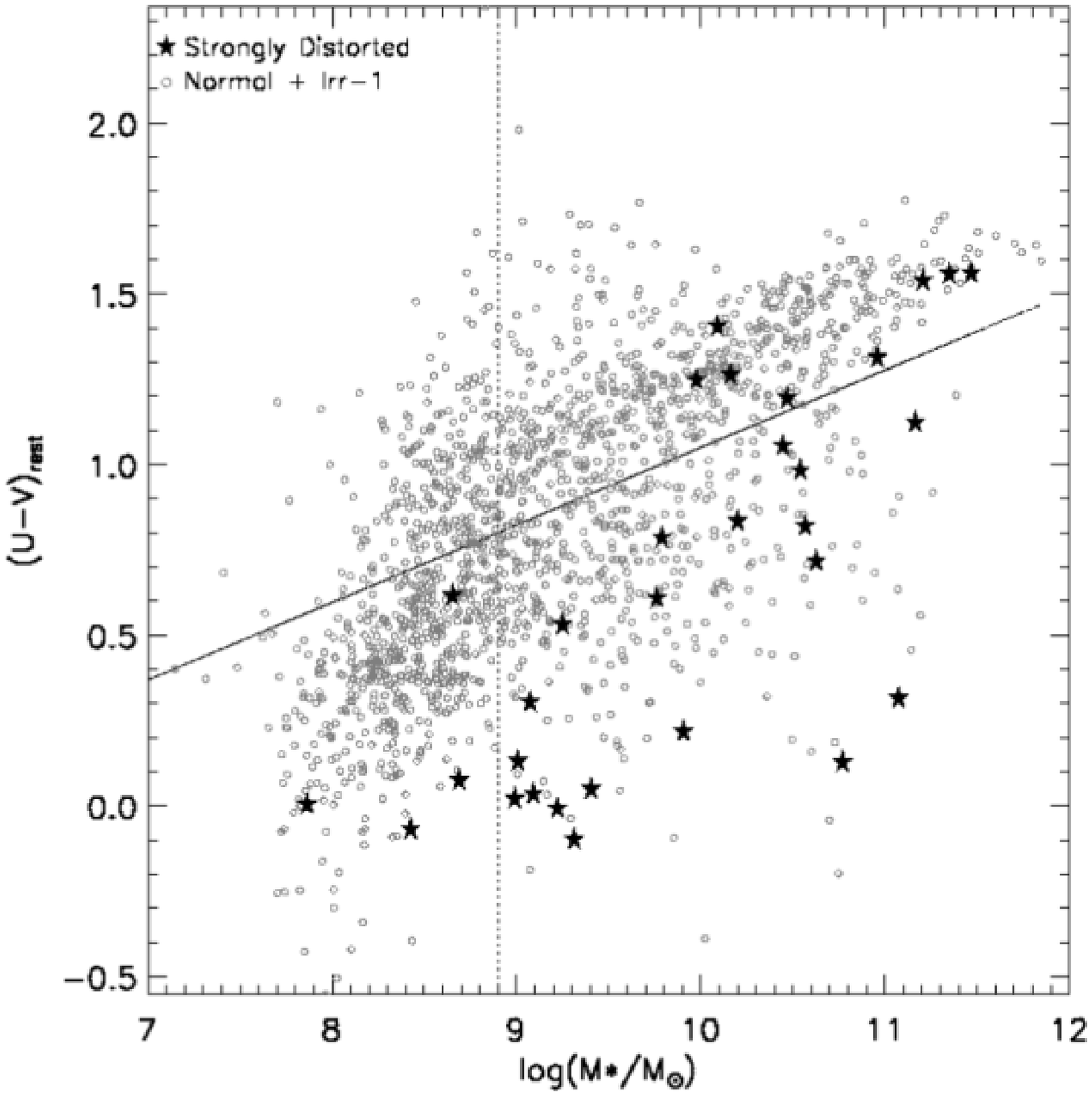}
\vspace{-0.4cm}
\caption{{\it left:} Strongly distorted galaxies (crosses) are
shown on the ICM density (greyscale) and DM surface mass density
$\kappa$ (contours range from 1.5$\sigma$ to 7.5$\sigma$ in intervals
of 1.5, where $\sigma\sim$0.03).  Most strongly distorted
galaxies lie outside region of high ICM density and DM peaks ($\kappa
\sim 0.1$).  {\it right:} Strongly distorted (black stars)
and normal relatively undisturbed galaxies (dots) are shown on
the color {\it versus} stellar mass plane. The vertical dotted line
denotes the COMBO-17 mass completeness limit for the red sequence.
Most (74\%) of the strongly distorted galaxies lie on the blue cloud
compared to 26\% of the red sequence.}
\end{figure}

\end{document}